\documentclass[journal,11pt,onecolumn,draftclsnofoot]{IEEEtran}
\ifCLASSINFOpdf
\else
\fi
\usepackage{amsfonts}
\usepackage{graphicx}
\usepackage{subfigure}
\usepackage{color}
\usepackage{amssymb}
\usepackage{amsmath}
\usepackage{amsbsy}
\usepackage{indentfirst}
\usepackage{cite}

\newtheorem{theorem}{Theorem}

\newtheorem{definition}{Definition}
\newtheorem{proposition}{Proposition}
\newtheorem{lemma}{Lemma}

\usepackage[font=small]{caption}

\begin{document}
%
\title{ On $q$-ratio CMSV for sparse recovery}
%
%
%

\author{Zhiyong~Zhou 
        and~Jun~Yu 
\thanks{The authors are with the Department
of Mathematics and Mathematical Statistics, Ume{\aa} University,  Ume{\aa},
901 87, Sweden (e-mail: zhiyong.zhou@umu.se, jun.yu@umu.se).}}
\maketitle

\begin{abstract}
Sparse recovery aims to reconstruct an unknown spare or approximately sparse signal from significantly few noisy incoherent linear measurements. As a kind of computable incoherence measure of the measurement matrix, $q$-ratio constrained minimal singular values (CMSV) was proposed in Zhou and Yu \cite{zhou2018sparse} to derive the performance bounds for sparse recovery. In this paper, we study the geometrical property of the $q$-ratio CMSV, based on which we establish new sufficient conditions for signal recovery involving both sparsity defect and measurement error. The $\ell_1$-truncated set $q$-width of the measurement matrix is developed as the geometrical characterization of $q$-ratio CMSV. In addition, we show that the $q$-ratio CMSVs of a class of structured random matrices are bounded away from zero with high probability as long as the number of measurements is large enough, therefore satisfy those established sufficient conditions. Overall, our results generalize the results in Zhang and Cheng \cite{zc} from $q=2$ to any $q\in(1,\infty]$ and complement the arguments of $q$-ratio CMSV from a geometrical view.
\end{abstract}


%
\IEEEpeerreviewmaketitle

\section{Introduction}
%
%
%
%

\IEEEPARstart{S}{parse} recovery (for instance, compressive sensing, introduced in \cite{crt,d}) concerns the reconstruction of a sparse or compressible (approximately sparse) signal $x\in\mathbb{R}^N$ from noisy underdetermined linear measurements: \begin{align}
y=Ax+e,
\end{align}
where $y\in\mathbb{R}^m$ is the measurement vector with $m\ll N$, $A\in\mathbb{R}^{m\times N}$ is the measurement matrix, and $e\in\mathbb{R}^m$ is the noise vector. As is well-known that if the measurement matrix $A$ satisfies the null space property (NSP) or restricted isometry property (RIP), then stable (w.r.t. sparsity defect) and robust (w.r.t. measurement error) recovery can be guaranteed via various algorithms, see \cite{fr,ek} for a comprehensive view.

However, due to the hardness of verifying NSP and computing restricted isometry constant (RIC) (see \cite{bdms} and \cite{tp}), new kinds of computable quality measures of the measurement matrices are proposed. Specifically, the authors in \cite{tn1} developed $\ell_1$-constrained minimal singular values (CMSV) $\rho_s(A)=\min\limits_{z\neq 0, \lVert z\rVert_1^2/\lVert z\rVert_2^2\leq s}\frac{\lVert Az\rVert_2}{\lVert z\rVert_2}$ and derived the $\ell_2$-norm recovery error bounds. Its geometrical property was investigated in \cite{zc}. An equivalent quantity called strong restricted eigenvalue was also used in \cite{bellec2016slope} and \cite{derumigny2018improved}. Meanwhile, in \cite{tn3}, the authors defined another similar quantity $\omega_{\lozenge}(A,s)=\min\limits_{z\neq 0, \lVert z\rVert_1/\lVert z\rVert_\infty\leq s}\frac{\lVert Az\rVert_{\lozenge}}{\lVert z\rVert_\infty}$ with $\lVert\cdot\rVert_{\lozenge}$ denoting a general norm, and obtained the performance bounds on the $\ell_{\infty}$ norm of the reconstruction errors. Their corresponding generalizations to block sparse recovery and low-rank matrix recovery were done in \cite{tn4} and \cite{tn2}. Very recently, \cite{zhou2018sparse} generalized these two quantities to a more general quantity called $q$-ratio CMSV with any $q\in(1,\infty]$, and established the performance bounds for both $\ell_q$ norm and $\ell_1$ norm of the reconstruction errors. 

In this paper, we study the geometrical property of $q$-ratio CMSV and make the following contributions. First, we define the $\ell_1$-truncated set $q$-width and connect it with the $q$-ratio CMSV. Second, by using a newly developed geometrical tool, we establish the sufficient conditions and both $\ell_q$ and $\ell_1$ norm error bounds for the recovery of both exactly sparse and compressible signals via Basis Pursuit (BP) \cite{cds} algorithm. Third, we show that a class of structured random matrices satisfy these sufficient conditions, since their $q$-ratio CMSVs are bounded away from zero with high probability as long as the number of measurements is large enough.

Throughout the paper, for any vector $z\in\mathbb{R}^N$ with entries $z_i$, we denote the $\ell_0$ norm $\lVert z\rVert_0=\sum_{i=1}^N 1\{z_i\neq 0\}$, the $\ell_{\infty}$ norm $\lVert z\rVert_{\infty}=\max\limits_{1\leq i\leq N}|z_i|$ and the $\ell_q$ norm $\lVert z\rVert_{q}=(\sum_{i=1}^N |z_i|^q)^{1/q}$ for $q\in(0,\infty)$. A signal $x$ is called $k$-sparse if $\lVert x\rVert_0\leq k$. We let $[N]$ be the set $\{1,2,\cdots,N\}$ and $|S|$ be the cardinality of a set $S$.  We write $S^c$ for the complement $[N]\setminus S$ of a set $S$ in $[N]$. For a vector $x\in\mathbb{R}^N$ and a set $S\subset [N]$, we denote by $x_S$ the vector coincides with $x$ on the indices in $S$ and is extended to zero outside $S$. For any matrix $A\in\mathbb{R}^{m\times N}$, $Ker A:=\{z\in\mathbb{R}^N: Az=0\}$. The $q$-radius of a set $T\subset \mathbb{R}^N$ is defined as $rad_q(T)=\sup_{t\in T}\lVert t\rVert_q$. We denote the unit ball of the $\ell_q$-norm in $\mathbb{R}^N$ by $B_q^N$ and its unit sphere by $S_q^{N-1}$. $\langle\cdot,\cdot\rangle$ denotes the inner product function in $\mathbb{R}^N$.

\section{$q$-ratio Sparsity, $q$-ratio CMSV and $\ell_1$-truncated set $q$-width}

We start with a kind of effective sparsity measure, called $q$-ratio sparsity, based on which the $q$-ratio CMSV is developed. It is entropy-based and has many nice properties, see \cite{l1,l2} for more detailed arguments. \\

\begin{definition}(\cite{l1,l2})
	For any non-zero $z\in\mathbb{R}^N$ and non-negative $q\notin\{0,1,\infty\}$, the $q$-ratio sparsity level of $z$ is defined as \begin{align}
	s_{q}(z)=\left(\frac{\lVert z\rVert_1}{\lVert z\rVert_q}\right)^{\frac{q}{q-1}}.
	\end{align} The cases of $q\in\{0,1,\infty\}$ are evaluated as limits: 
	$s_0(z)=\lim\limits_{q\rightarrow 0} s_q(z)=\lVert z\rVert_0$, $s_1(z)=\lim\limits_{q\rightarrow 1} s_q(z)=\exp(H_1(\pi(z)))$, $s_\infty(z)=\lim\limits_{q\rightarrow 0} s_q(z)=\frac{\lVert z\rVert_1}{\lVert z \rVert_\infty}$.
	Here $\pi(z)\in\mathbb{R}^N$ with entries $\pi_i(z)=|z_i|/\lVert z\rVert_1$ and $H_1$ is the ordinary Shannon entropy $H_1(\pi(z))=-\sum_{i=1}^N \pi_i(z)\log \pi_i(z)$.\\
\end{definition}

Now, we are ready to present the definition of $q$-ratio CMSV, which is a computable quality measure for the measurement matrix. The related theoretical sparse recovery results and computational aspects are referred to \cite{zhou2018sparse}. In this paper, we investigate it geometrically instead.\\

\begin{definition}(\cite{zhou2018sparse})
	For any real number $s\in[1,N]$, $q\in(1,\infty]$ and matrix $A\in\mathbb{R}^{m\times N}$, the $q$-ratio constrained minimal singular value (CMSV) of $A$ is defined as \begin{align}
	\rho_{q,s}(A)=\min\limits_{z\neq 0,s_q(z)\leq s}\,\,\frac{\lVert Az\rVert_2}{\lVert z\rVert_q}. 
	\end{align} 
\end{definition}

\medskip
To exploit the geometrical property of $q$-ratio CMSV, we define the $\ell_1$-truncated set $q$-width of a measurement matrix $A$ as follows. \\

\begin{definition}
	For any $r>0$, $q\in (1,\infty]$ and matrix $A\in\mathbb{R}^{m\times N}$, we define the $\ell_1$-truncated set $q$-width of $A$ as \begin{align}
	R_{q,r}(A)=\min\limits_{u\in B_1^N \cap\,r S_q^{N-1}}\lVert Au\rVert_2. 
	\end{align}
\end{definition}

\medskip
In what follows, the proposition gives a geometrical characterization of $q$-ratio CMSV. \\

\begin{proposition}
	The following equation connects the $q$-ratio CMSV and the $\ell_1$-truncated set $q$-width: \begin{align}
	\rho_{q,r^{-\frac{q}{q-1}}}(A)=r^{-1}\cdot R_{q,r}(A). 
	\end{align} 
\end{proposition}

\noindent 
{\bf Proof.} Based on the definitions, we have \begin{align*}
\rho_{q,r^{-\frac{q}{q-1}}}(A)&=\min\limits_{z\neq 0, s_q(z)\leq r^{-\frac{q}{q-1}}} \frac{\lVert Az\rVert_2}{\lVert z\rVert_q} \\
&=\min\limits_{z\neq 0, \left\lVert \frac{z}{\lVert z\rVert_q}\right\rVert_1^{\frac{q}{q-1}}\leq r^{-\frac{q}{q-1}}} \left\lVert A\frac{z}{\lVert z\rVert_q}\right\rVert_2 \\
&=\min\limits_{\lVert v\rVert_q=1,\lVert v\rVert_1^{\frac{q}{q-1}}\leq r^{-\frac{q}{q-1}}} \lVert Av\rVert_2 \\
&=\min\limits_{\lVert u\rVert_q=r,\lVert u\rVert_1\leq 1} r^{-1}\lVert Au\rVert_2 \\
&=r^{-1}\cdot R_{q,r}(A).
\end{align*}

The following proposition is used to exploit the geometrical property of $q$-ratio CMSV, which is an extension of Lemma 1 in \cite{zc} and Proposition 5.1.2 in \cite{chafai2012interactions}. Their results correspond to the special case of $\alpha=0$. \\

\begin{proposition}
	Let $T$ be a star body with respect to the origin that is a compact subset $T$ of $\mathbb{R}^N$ such that for any $t\in T$, the segment between 0 and $t$ is contained in $T$, and for any $\alpha\geq 0$, $T_0=\{h: \lVert \Phi h\rVert_2\leq \alpha\}$, where $\Phi$ is an $m\times N$ matrix with row vectors denoted by $Y_1, Y_2,\cdots, Y_m$. If \begin{align}
	\min\limits_{u\in T\cap\,r S_q^{N-1}} \sum\limits_{i=1}^m \langle Y_i, u\rangle^2>\alpha^2,
	\end{align}
	then $rad_q(T_0 \cap T)<r$.\\
\end{proposition}

\noindent
{\bf Proof.} Select any vector $t\in T$, and assume that $\lVert t\rVert_q\geq r$. Then $z=\frac{r}{\lVert t\rVert_q}t\in T\cap rS_q^{N-1}$, which implies that $\rVert \Phi z\rVert_2^2=\sum\limits_{i=1}^m \langle Y_i, z\rangle^2>\alpha^2$. Therefore, $z\notin T_0 \Rightarrow t=\frac{\lVert t\rVert_q}{r}z\notin T_0$. Hence, by using the contrapositive conclusion, for any $t\in T_0 \cap T$, it holds that $\lVert t\rVert_q< r$. In other words, we have 
$rad_q(T_0 \cap T)=\sup\limits_{t\in T_0\cap T}\lVert t\rVert_q<r$, which completes the proof.\\

By using Proposition 2 and combining the facts that the unit ball of $\ell_1$ norm $B_1^N$ is a star body and that \begin{align*}
R_{q,r}^2(A)=\min\limits_{u\in B_1^N\cap rS_q^{N-1}} \lVert Au\rVert_2^2=\min\limits_{u\in B_1^N\cap rS_q^{N-1}}\sum\limits_{i=1}^m \langle a_i, u\rangle^2,
\end{align*}
we obtain the following proposition immediately. As shown later, it plays a crucial role in deriving the robust recovery results involving the measurement error via noisy BP algorithm, which was not considered in \cite{zc,chafai2012interactions}.\\

\begin{proposition}
	If $R_{q,r}(A)>\alpha$, then $rad_q(T_0\cap B_1^N)<r$. In particular, taking $\alpha=0$, we have if $R_{q,r}(A)>0$, then $rad_q(Ker A\cap B_1^N)<r$. 
\end{proposition}

\section{Sparse Recovery results}

In this section, we first study the recovery conditions for the noise free BP algorithm: \begin{align}
\min\limits_{z\in\mathbb{R}^N} \lVert z\rVert_1,\,\ \text{s.t.}\,\,Az=y, \label{bp}
\end{align}
where $y=Ax$. First, it is well-known that when the true signal $x$ is $k$-sparse, the sufficient and necessary condition for the exact recovery of the noise free BP problem is given by the NSP of order $k$:
\begin{align*}
\lVert h_S\rVert_1<\lVert h_{S^c}\rVert_1, \forall\,\,h\in Ker A\setminus \{0\}, S\subset [N]\,\text{with}\,|S|\leq k,
\end{align*}
see Theorem 4.5 in \cite{fr}. Then, a sufficient condition for exact recovery of $k$-sparse signal via noise free BP (\ref{bp}), which generalizes Lemma 2 in \cite{zc} and Proposition 5.1.1 in \cite{chafai2012interactions} from $q=2$ to any $q\in(1,\infty]$, goes as follows. \\

\begin{proposition}
	If there exists some $q\in(1,\infty]$ such that $rad_q(Ker A \cap B_1^N)<r$ with $r\leq \frac{1}{2}k^{-\frac{q-1}{q}}$, then for every $k$-sparse signal $x$, the solution to the noise free BP algorithm (\ref{bp}) is unique and equal to $x$.\\
\end{proposition}

\noindent
{\bf Proof.} Let $h\in Ker A\setminus\{0\}$, and $S\subset [N]$ with $|S|\leq k$. By our assumption $rad_q(Ker A \cap B_1^N)<r\leq \frac{1}{2}k^{-\frac{q-1}{q}}$, we have \begin{align*}
\lVert h\rVert_q<r\lVert h\rVert_1\leq \frac{1}{2}k^{-\frac{q-1}{q}}\lVert h\rVert_1.
\end{align*}
Thus, \begin{align*}
\lVert h_S\rVert_1\leq k^{1-1/q}\lVert h_S\rVert_q&\leq k^{1-1/q}\lVert h\rVert_q \\
&<k^{1-1/q}\cdot \frac{1}{2}k^{-\frac{q-1}{q}}\lVert h\rVert_1=\frac{1}{2}\lVert h\rVert_1,
\end{align*}
which leads to $\lVert h_S\rVert_1<\lVert h_{S^c}\rVert_1$. The proof is completed by using the NSP of order $k$. \\

Then a sufficient condition based on $q$-ratio CMSV is derived as follows.\\

\begin{theorem}(Exactly sparse recovery)
	If there exists some $q\in (1,\infty]$ such that $\rho_{q,s}(A)>0$ with $s\geq 2^{\frac{q}{q-1}} k$, then for every $k$-sparse signal $x$, the solution to the noise free BP algorithm (\ref{bp}) is unique and equal to $x$. \\
\end{theorem}

\noindent
{\bf Proof.} By using Proposition 1 and the assumption, we have \begin{align}
\rho_{q,s}(A)=r^{-1}\cdot R_{q,r}(A)>0,
\end{align} 
and $s=r^{-\frac{q}{q-1}}\geq 2^{\frac{q}{q-1}}k$. Thus $R_{q,r}(A)>0$ and $r\leq (2^{\frac{q}{q-1}} k)^{-\frac{q-1}{q}}=\frac{1}{2}k^{-\frac{q-1}{q}}$. By Proposition 3, we obtain that $rad_q(Ker A\cap B_1^N)<r$ with $r\leq \frac{1}{2}k^{-\frac{q-1}{q}}$. Then Proposition 4 implies the desired result. \\

\noindent
{\bf Remark.} The condition that $\rho_{q,s}(A)>0$ with $s\geq 2^{\frac{q}{q-1}} k$ is equivalent to the condition $\rho_{q,2^{\frac{q}{q-1}}k}(A)>0$ by using the non-increasing property of $\rho_{q,s}(A)$ with respect to $s$. Actually this sufficient condition can also obtained as a by product of the error bound results (11) and (12) in \cite{zhou2018sparse} for exactly sparse recovery via noisy BP.\\

Now, we are ready to establish the recovery result for compressible signals via noise free BP, with Proposition 5 in \cite{zc} corresponding to the special case of $q=2$. In the following context, we let the $\ell_1$-error of best $k$-term approximation of $x$ be $\sigma_k(x)_1=\inf \{\lVert x-z\rVert_1, z\in\mathbb{R}^N \text{is $k$-sparse}\}$ and assume that $S$ is the index set that contains the largest $k$ absolute entries of $x$ so that $\sigma_k(x)_1=\lVert x_{S^c}\rVert_1$.\\

\begin{theorem}(Stable recovery) Let $\hat{x}$ be a minimizer of the noise free BP (\ref{bp}). For any $1<q\leq \infty$, assume that $\rho_{q,s}(A)>0$ with $s\geq 4^{\frac{q}{q-1}}k$, then \begin{align}
	\lVert \hat{x}-x\rVert_1&\leq 4\sigma_k(x)_1, \\ 
	\lVert \hat{x}-x\rVert_q&\leq k^{1/q-1}\sigma_k(x)_1. \label{c2} 
	\end{align}	
\end{theorem}

\noindent
{\bf Proof.}  As discussed in the Proof of Theorem 1, the assumption $\rho_{q,s}(A)>0$ with $s\geq 4^{\frac{q}{q-1}}k$ implies that $rad_q(Ker A\cap B_1^N)<r$ with $r\leq \frac{1}{4}k^{-\frac{q-1}{q}}$. If $h\in Ker A\setminus \{0\}$, then we have \begin{align}
\lVert h_S\rVert_1\leq k^{1-1/q}\lVert h_S\rVert_q&\leq k^{1-1/q}\lVert h\rVert_q \nonumber \\
&<k^{1-1/q}r\lVert h\rVert_1\leq \frac{1}{4}\lVert h\rVert_1,
\end{align}
Therefore, it holds that $\lVert h_S\rVert_1< \frac{1}{3}\lVert h_{S^c}\rVert_1$ for $h\in Ker A\setminus \{0\}$. Then obviously we can obtain \begin{align}
\lVert h_S\rVert_1\leq  \frac{1}{3}\lVert h_{S^c}\rVert_1,\,\,\text{for all $h\in KerA$}.  \label{stablensp}
\end{align}

On the other hand, since $\hat{x}$ is a minimizer of the noise free BP (\ref{bp}), by taking $h=\hat{x}-x$, we have \begin{align*}
\lVert x\rVert_1\geq \lVert \hat{x}\rVert_1&=\lVert x+h\rVert_1 \\
&=\lVert x_S+h_S\rVert_1+\lVert x_{S^c}+h_{S^c}\rVert_1\\
&\geq \lVert x_S\rVert_1-\lVert h_S\rVert_1-\lVert x_{S^c}\rVert_1+\lVert h_{S^c}\rVert_1, \label{err}
\end{align*}
which implies that \begin{align}
\lVert h_{S^c}\rVert_1\leq \lVert h_S\rVert_1+2\lVert x_{S^c}\rVert_1 
\end{align}
Since $h=\hat{x}-x\in Ker A$, then combining (\ref{stablensp}) and (\ref{err}) leads to $\lVert h_{S^c}\rVert_1\leq 3\lVert x_{S^c}\rVert_1$. As a consequence, \begin{align}
\lVert \hat{x}-x\rVert_1=\lVert h\rVert_1&=\lVert h_S\rVert_1+\lVert h_{S^c}\rVert_1 \nonumber \\
&\leq (1+\frac{1}{3})\lVert h_{S^c}\rVert_1 \nonumber \\
&\leq 4\lVert x_{S^c}\rVert_1=4\sigma_k(x)_1.
\end{align}

As for (\ref{c2}), if $h=\hat{x}-x\equiv0$, then it holds trivially. And if $h\in Ker A\setminus \{0\}$, we have \begin{align}
\lVert h\rVert_q<r\lVert h\rVert_1\leq \frac{1}{4}k^{-\frac{q-1}{q}}\cdot 4\sigma_k(x)_1=k^{1/q-1}\sigma_k(x)_1.
\end{align}
The proof is completed. \\

Next, we extend the result to the noisy BP case, which has not been considered yet in the existing literature \cite{zc,chafai2012interactions}. That is, we focus on the following problem:  \begin{align}
\min\limits_{z\in\mathbb{R}^N} \lVert z\rVert_1,\,\ \text{s.t.}\,\,\lVert Az-y\rVert_2\leq \varepsilon, \label{nbp}
\end{align}
where $y=Ax+e$ with $\lVert e\rVert_2\leq \varepsilon$. The key gradient for the proof is the newly developed Proposition 3 in Section II. \\

\begin{theorem}(Stable and robust recovery) Let $\hat{x}$ be a minimizer of the noisy BP (\ref{nbp}). For any $1<q\leq \infty$, assume that $\rho_{q,s}(A)>0$ with $s\geq 4^{\frac{q}{q-1}}k$, then \begin{align}
	\lVert \hat{x}-x\rVert_1&\leq 4\sigma_k(x)_1+\frac{8\varepsilon}{\rho_{q,s}(A)}k^{1-\frac{1}{q}},  \label{c3} \\ 
	\lVert \hat{x}-x\rVert_q&\leq k^{1/q-1}\sigma_k(x)_1+\frac{2\varepsilon}{\rho_{q,s}(A)}. \label{c4} 
	\end{align}	
\end{theorem}

\noindent
{\bf Proof.} We first show (\ref{c3}). We assume $h=\hat{x}-x\neq 0$ and $\lVert h\rVert_1>\frac{8\varepsilon}{\rho_{q,s}(A)}k^{1-\frac{1}{q}}$ with $s\geq 4^{\frac{q}{q-1}}k$, otherwise (\ref{c3}) holds trivially. Then we have $\rho_{q,s}(A)>\frac{8\varepsilon k^{1-\frac{1}{q}}}{\lVert h\rVert_1}$. By using Proposition 1 and the assumption, we have $\rho_{q,s}(A)=r^{-1}\cdot R_{q,r}(A)>\frac{8\varepsilon k^{1-\frac{1}{q}}}{\lVert h\rVert_1}$ and $s=r^{-\frac{q}{q-1}}\geq 4^{\frac{q}{q-1}}k$. Thus $R_{q,r}(A)>\frac{8r\varepsilon k^{1-\frac{1}{q}}}{\lVert h\rVert_1}$ and $r\leq (4^{\frac{q}{q-1}}k)^{-\frac{q-1}{q}}=\frac{1}{4}k^{\frac{1}{q}-1}$. As a consequence, Proposition 3 implies that $rad_{q}(T_0\cap B_1^N)<r$ with $T_0=\{z:\lVert Az\rVert_2\leq \alpha=\frac{8r\varepsilon k^{1-\frac{1}{q}}}{\lVert h\rVert_1}\}$. And it is easy to verify that $\frac{4rk^{1-\frac{1}{q}}}{\lVert h\rVert_1}h\in T_0\cap B_1^N$ by using the facts that $\lVert Ah\rVert_2=\lVert (A\hat{x}-y)-(Ax-y)\rVert_2\leq \lVert  A\hat{x}-y\rVert_2+\lVert e\rVert_2\leq 2\varepsilon$ and  $r\leq \frac{1}{4}k^{\frac{1}{q}-1}$. Therefore, we obtain \begin{align*}
\left\lVert \frac{4rk^{1-\frac{1}{q}}}{\lVert h\rVert_1}h\right\rVert_q<r\Rightarrow \lVert h\rVert_q<\frac{1}{4}k^{\frac{1}{q}-1}\lVert h\rVert_1.
\end{align*}

Furthermore, since $\hat{x}$ is a minimizer of the noisy BP (\ref{nbp}), by taking $h=\hat{x}-x$ and using the same argument as that used to obtain (\ref{err}), we still have \begin{align}
\lVert h_{S^c}\rVert_1\leq \lVert h_S\rVert_1+2\lVert x_{S^c}\rVert_1.
\end{align}
Hence, \begin{align}
\lVert \hat{x}-x\rVert_1&=\lVert h\rVert_1=\lVert h_S\rVert_1+\lVert h_{S^c}\rVert_1 \nonumber \\
&\leq 2\lVert h_S\rVert_1+2\lVert x_{S^c}\rVert_1 \nonumber \\
&\leq 2k^{1-\frac{1}{q}}\lVert h\rVert_q+2\lVert x_{S^c}\rVert_1  \label{inequality} \\
&<\frac{1}{2}\lVert h\rVert_1+2\lVert x_{S^c}\rVert_1 \nonumber \\
\Rightarrow \lVert h\rVert_1&<4\lVert x_{S^c} \rVert_1=4\sigma_k(x)_1, \nonumber 
\end{align}
which completes the proof of (\ref{c3}). 

To show (\ref{c4}), we assume that $h=\hat{x}-x\neq 0$ and $\lVert h\rVert_q>\frac{2\varepsilon}{\rho_{q,s}(A)}$ with $s\geq 4^{\frac{q}{q-1}}k$, otherwise (\ref{c4}) holds trivially. We first show that under this assumption we have  \begin{align}
\lVert h\rVert_q<\frac{1}{4}k^{\frac{1}{q}-1}\lVert h\rVert_1\,\,\text{for $h=\hat{x}-x\neq 0$.} 
\end{align}
We prove it by contradiction. If for $h=\hat{x}-x\neq 0$, $\lVert h\rVert_q\geq \frac{1}{4}k^{\frac{1}{q}-1}\lVert h\rVert_1$, i.e., $s_{q}(h)\leq 4^{\frac{q}{q-1}}k$, then, based on the definition, we have \begin{align}
\rho_{q,s}(A)=\min\limits_{z\neq 0, s_q(z)\leq s}\frac{\lVert Az\rVert_2}{\lVert z\rVert_q}&\leq \min\limits_{z\neq 0, s_q(z)\leq 4^{\frac{q}{q-1}}k}\frac{\lVert Az\rVert_2}{\lVert z\rVert_q}
\nonumber \\
&\leq \frac{\lVert Ah\rVert_2}{\lVert h\rVert_q}\leq \frac{2\varepsilon}{\lVert h\rVert_q} \nonumber \\
\Rightarrow \lVert h\rVert_q&\leq \frac{2\varepsilon}{\rho_{q,s}(A)}. 
\end{align}
This is a contradiction with our assumption. Thus we have $\lVert h\rVert_1\geq 4k^{1-\frac{1}{q}}\lVert h\rVert_q>\frac{8\varepsilon}{\rho_{q,s}(A)}k^{1-\frac{1}{q}}$ with $s\geq 4^{\frac{q}{q-1}}k$. Finally, by adopting the result in proving (\ref{c3}), we can obtain that \begin{align}
\lVert h\rVert_q\leq \frac{1}{4}k^{\frac{1}{q}-1}\lVert h\rVert_1<\frac{1}{4}k^{\frac{1}{q}-1}\cdot 4\sigma_k(x)_1=k^{\frac{1}{q}-1}\sigma_k(x)_1,
\end{align}
which ends the proof of (\ref{c4}). \\

\noindent
{\bf Remark.} If we go back by combining (\ref{c4}) and (\ref{inequality}), we can improve the error bound (\ref{c3}) to \begin{align}\lVert \hat{x}-x\rVert_1&\leq 4\sigma_k(x)_1+\frac{4\varepsilon}{\rho_{q,s}(A)}k^{1-\frac{1}{q}}. \label{c5}
\end{align}
Then (\ref{c4}) and (\ref{c5}) are exactly the results (17) and (18) given in \cite{zhou2018sparse}, while we use a different proof procedure based on the geometrical characterization of $q$-ratio CMSV. 

\section{Structured Random Matrices}
The theoretical results of $q$-ratio CMSV for subgaussian random matrices were established in \cite{zhou2018sparse}. In this section, we further study the $q$-ratio CMSV for a class of structured random matrices. More specifically, given an $N\times N$ orthogonal matrix. The measurement matrix is generated by selecting $m$ rows of this matrix at random. Consider a system of vectors $\phi_1,\cdots,\phi_n$ such that they are orthogonal, and for $1\leq i\leq N$, $\lVert \phi_i\rVert_\infty\leq 1/\sqrt{N}$ and $\lVert \phi_i\rVert_2=M$ where $M$ is a fixed number. Define the random vector $a$ as $a=\phi_i$ with probability $1/N$ and let $a_1,\cdots,a_N$ be independent copies of $a$. Then the $m\times N$ random measurement matrix $A$ is constructed by taking its $i$-th row to be $a_i$. This kind of structured random matrices include the class of Fourier random matrices and the class of Hadamard random matrices. Next, we start with two key existing lemmas.\\

\begin{lemma}(Proposition 6 in \cite{zc})
	Let $m\times N$ random matrix $A$ be defined as above with the number of measurements satisfying \begin{align}
	m\geq \frac{9\delta^2 C^2 (\log m)^3\log N}{M^2}\cdot k.
	\end{align}
	Then with probability greater than $1-C\exp(-cM^2ms^{-1})$, it has $\rho_{2,s}(A)>0$ with $s\geq \delta^2 k$, and guarantees exactly sparse signals recovery with $\delta=2$ and stable (and robust) recovery with $\delta=4$, where $C,c$ are some constants. \\
\end{lemma}

\begin{lemma}(Proposition 2 in \cite{zhou2018sparse})
	If $1<q_2\leq q_1\leq\infty$, then for any real number $1\leq s\leq N^{\frac{q_1(q_2-1)}{q_2(q_1-1)}}$, we have \begin{align*}
	\rho_{q_1,s}(A)\geq \rho_{q_2,s^{\frac{q_2(q_1-1)}{q_1(q_2-1)}}}(A)\geq  s^{-\frac{q_2(q_1-1)}{q_1(q_2-1)}} \rho_{q_1, s^{\frac{q_2(q_1-1)}{q_1(q_2-1)}}}(A). \label{rhoineq}
	\end{align*}
	In particular, we have if $1<q<2$, then $\rho_{q,s}(A)\geq s^{-1}\rho_{2,s}(A)$, while if $2\leq q\leq \infty$, then $\rho_{q,s}(A)\geq \rho_{2,s^{\frac{2(q-1)}{q}}}(A)$.\\
\end{lemma}

We see that the common assignment for successful recovery is to choose $A$ such that $\rho_{q,s}(A)>0$ with $s\geq \delta^{\frac{q}{q-1}} k$. Then the following probabilistic statements about $q$-ratio CMSV holds immediately by combining Lemma 1 and Lemma 2. \\

\begin{theorem}
	Let $m\times N$ random matrix $A$ be defined as before. \\
	\noindent	
	(a) When $1<q<2$, if the number of measurements satisfying \begin{align}
	m\geq \frac{9\delta^{\frac{q}{q-1}} C^2 (\log m)^3\log N}{M^2}\cdot k,
	\end{align}
	then with probability greater than $1-C\exp(-cM^2ms^{-1})$, it has $\rho_{q,s}(A)>0$ with $s\geq \delta^{\frac{q}{q-1}} k$, and guarantees exactly sparse signals recovery with $\delta=2$ and stable (and robust) recovery with $\delta=4$, where $C,c$ are some constants used in Lemma 1.  
	
	\noindent	
	(b) When $2\leq q\leq\infty$, if the number of measurements satisfying \begin{align}
	m\geq \frac{9\delta^{2} C^2 (\log m)^3\log N}{M^2}\cdot k^{\frac{2(q-1)}{q}},
	\end{align}
	then with probability greater than $1-C\exp(-cM^2ms^{-1})$, it has $\rho_{q,s}(A)>0$ with $s\geq \delta^{\frac{q}{q-1}} k$, and guarantees exactly sparse signals recovery with $\delta=2$ and stable (and robust) recovery with $\delta=4$, where $C,c$ are some constants used in Lemma 1.  \\
	
\end{theorem}

\section{Conclusion}
In this paper, we studied the $q$-ratio CMSV geometrically by introducing the $\ell_1$-truncated set $q$-width. The geometrical property of $q$-ratio CMSV was used to derive sufficient conditions and error bounds for both exactly sparse and compressible signal recovery. A class of structured random matrices is shown to satisfy these $q$-ratio CMSV based sufficient conditions.

\section*{Acknowledgment}
This work is supported by the Swedish Research Council grant (Reg.No. 340-2013-5342).

\ifCLASSOPTIONcaptionsoff
  \newpage
\fi



%
\bibliographystyle{IEEEtran}
\end{document}